# Идентификация объектов оптического диапазона в областях жесткого излучения вблизи красных карликовых звёзд

*А.А. Шляпников*

Федеральное государственное бюджетное учреждение науки «Крымская астрофизическая обсерватория РАН»

*Аннотация*

Проведен анализ областей локализации источников ТэВ гамма-излучения в рентгеновском и оптическом диапазонах спектра. Указаны расстояния от положения максимумов в распределении высокоэнергетических потоков, до вероятных кандидатов на идентификацию с красными карликами. Также рассмотрены возможные отождествления более слабых ТэВ источников и других объектов поля.

Ключевые слова: рентгеновские и гамма-источники, звезды - переменные и пекулярные.

aas@craocrimea.ru

*Введение*

Проблема идентификации первых «дискретных» источников жесткого диапазона спектра ($10^{-8} - 10^{-11}$ эрг ≈ 0.6 МэВ – 0.6 кэВ) в оптике была исторически связана с невысокой точностью в определении их координат. Так первый дискретный рентгеновский источник, был зафиксирован в 1962 г. в созвездии Скорпиона детектором, установленным на ракете, и получил обозначение Sco X-1 (Джиакони и др., 1962). После уточнения его координат (Гурский и др., 1966), он был с точностью около 1' идентифицирован со звездой V≈13$^m$, имеющей ультрафиолетовый избыток в 1966 г. (Сандадж и др., 1966). С момента открытия дискретного рентгеновского источника до его идентификации в оптическом диапазоне спектра прошло четыре года.

Более драматичной оказалась история идентификация источников гамма диапазона спектра. Наиболее характерным примером является поиск в других спектральных диапазонах источников гамма вспышек (GRB - Gamma Ray Burst).

Впервые информация о наблюдениях 16-ти гамма вспышек была опубликована в 1973 г. (Клибсейдол и др., 1973). Они регистрировались четырьмя спутниками серии Vela[1] в период с июля 1969 г. по июль 1972 г. в диапазоне энергий 0.2 – 1.5 МэВ, имели продолжительность от 0.1 до более 30 секунд и интегрированную по времени плотность потока от ~ $10^{-5}$ до ~ $2 \times 10^{-4}$ эрг × с × см$^{-2}$. Направление прихода гамма вспышек не были связаны ни с Землей

---

[1] https://heasarc.gsfc.nasa.gov/docs/heasarc/missions/vela5a.html

(спутники регистрировали гамма излучение, обусловленное испытанием ядерного оружия), ни с Солнцем.

Лишь в феврале 1997 г. 28.123620 UT через 24 года спутник BeppoSAX[2] зарегистрировал мультипиковую GRB продолжительностью около 80 секунд. Изображение, полученное с помощью позиционно-чувствительной камеры гамма диапазона WFC (Wide Field Camera), позволило определить предварительные координаты вспышки R.A. = $5^h01^m57^s$ и Decl. = +11°46'.4 на эпоху 2000.0 с ошибкой 3' (Коста и др., 1997). Через 8 часов, благодаря наведению на область вспышки рентгеновского телескопа NFI (Narrow Field Instruments) удалось улучшить координаты GRB. А триангуляция наблюдений спутников BeppoSAX и Ulysses[3] позволила определить координаты, по которым было обнаружено оптическое послесвечение - оптический транзиент. Угасающий источник был найден при сравнении наблюдений, полученных на телескопах имени Гершеля[4] и имени Ньютона[5] (Гроот и др., 1997). Дальнейшие наблюдения на телескопе имени Хаббла[6] позволили обнаружить не только ослабевающее послесвечение, но и возможную галактику – «хозяйку» оптического транзиента (Сахю и др., 1997). Так начиналась «Эра послесвечений» гамма вспышек. Некоторые из послесвечений были идентифицированы как сверхновые, природа других остается неизвестной до сих пор.

Ситуация с идентификацией объектов жесткого диапазона в оптике кардинально изменилась после запуска космических обсерваторий CGRO, Chandra, eROSITA, Fermi, GRANAT, INTEGRAL, MAXI, ROSAT, Rossi-XTE, XMM-Newton и создания межпланетной наблюдательной сети IPN (InterPlanetary Network)[7]. Больше информации об упомянутых проектах можно найти по ссылке[8].

Несмотря на высокую точность локализации источников жесткого излучения, составляющую угловые минуты, далеко не все обнаруженные источники идентифицированы в оптическом диапазоне спектра. Это не позволяет детально изучить их природу.

Если в космических наблюдениях удалось повысить точность определения координат объектов жесткого излучения, то иначе складывается ситуация с наземными наблюдениями на телескопах, регистрирующих черенковское излучение, вызванное потоками гамма квантов в земной атмосфере. При покрытии области неба в несколько градусов, угловое разрешение черенковских телескопов составляет от 15' до 2º, что делает весьма проблематичным идентификацию в оптическом диапазоне спектра гамма источников.

В данной работе представлен анализ областей локализации шести источников ТэВ гамма-излучения в оптическом диапазоне спектра. Указаны расстояния от положения максимумов в распределении высокоэнергетических потоков, до вероятных кандидатов на идентификацию с красными карликами. Также рассмотрены возможные отождествления более слабых источников гамма излучения с другими объектами.

*Жесткое излучение Солнца и звёзд*

Спокойное Солнце является источником мягкого рентгеновского излучения, генерируемого в короне. При энергии квантов менее $10^{-9}$ эрг их поток резко убывает и практически не наблюдается в гамма диапазоне. Это было обнаружено во время первых экспериментальных и теоретических исследований жесткого излучения Солнца (Петерсон, Винклер, 1959). Во время солнечных вспышек энергии квантов могут меняться в пределах от сотен кэВ до десятков МэВ (Холт, Рамати, 1969). При этом важно отметить корреляцию по времени между жестким рентгеновским и микроволновым излучениями. Поиск радиоимпульсов, совпадающих с наблюдениями в жестком диапазоне спектра по времени (Бэрд и др., 1975) не дал результатов. Однако, был получен верхний предел на радиоизлучение, исключающий звёздные вспышки, как источник гамма-всплесков, при условии, что они подобны солнечным, но большие по масштабу (Кавалло, Джелли, 1975).

---

[2] https://heasarc.gsfc.nasa.gov/docs/sax/sax.html
[3] https://web.archive.org/web/20110806115817/http://ulysses.jpl.nasa.gov/
[4] http://www.ing.iac.es/astronomy/telescopes/wht/
[5] http://www.ing.iac.es/astronomy/telescopes/int/
[6] https://www.nasa.gov/mission_pages/hubble/about
[7] https://heasarc.gsfc.nasa.gov/docs/heasarc/missions/ipn.html
[8] https://heasarc.gsfc.nasa.gov/docs/

По мере совершенствования наблюдательного оборудования проблема идентификации жесткого излучения исходящего от звёзд, в частности звёзд – карликов нижней части Главной последовательности, получила своё развитие. Так, в каталоге GTSh10, из 5535 звёзд с активностью солнечного типа (Гершберг и др., 2011) у близко 2000 объектов обнаружено рентгеновское излучение. Большая часть из них демонстрирует оптические вспышки со значительными амплитудами.

Уникальное событие было зарегистрировано для красного карлика DG CVn (Д'Элия и др., 2014). 23 апреля 2014 г. в 21:07:08 UT телескоп BAT (Burst Alert Telescope) на борту космической обсерватории Swift зарегистрировал вспышку в диапазоне энергий 15-50 кэВ, которая продолжалась 64 секунды. Дальнейшие наблюдения в рентгене позволили уточнить координаты вспышки до 4".9 и идентифицировать объект со вспыхивающей звездой спектрального типа M4.0V DG CVn.

Актуальность исследований вспышечной активности красных карликов обусловлена открытием экзопланет у нескольких тысяч звёзд, немалая часть которых является звёздами с активностью солнечного типа (Горбачев и др., 2019). Более подробно данная проблема изложена в статье (Кабальеро-Гарсия и др., 2016).

*Наблюдения областей красных карликов в ТэВ диапазоне*

Очевидно первые наблюдения областей, содержащих красные карлики в ТэВ диапазоне с энергиями ~ $10^{12}$ эВ, были выполнены в Крымской астрофизической обсерватории в 1994 г. (Алексеев и др., 1995). При проведении параллельных наблюдений EV Lac в оптическом диапазоне на телескопе АЗТ-11 и на черенковском телескопе ГТ-48 в течение нескольких ночей, были обнаружены коррелирующие по времени события в оптике и гамма лучах. Наиболее статистически надежный результат в ТэВ диапазоне был получен незадолго до оптической вспышки 31 августа 1994 г. в 19:40 UT. Учитывая несовпадение координат вспышки и EV Lac на 0°.6, авторы сообщили о возможной связи гамма вспышки и вспышки в оптическом диапазоне.

В статьях «Галактические космические лучи: первое обнаружение ТэВ гамма-излучения от красных карликов» и «Новый тип источников галактических космических лучей» (Синицына и др., 2019, Синицына и др., 2020) представлены данные долгосрочных наблюдений в эксперименте SHALON, направленных на поиск излучения от звезд – активных красных карликов с интенсивностью выше 800 ГэВ. Все объекты расположены на угловых расстояниях менее 5° от наблюдаемых программных источников ТэВ-излучения (остатка сверхновой Тихо, Крабовидной туманности, объектов 4C+41.63 и Cyg X-3). Так как в эксперименте SHALON поле зрения телескопа больше 8°, то при наблюдении программных объектов в площадку одновременно попадают эруптивные (Er) и/или вспыхивающие (Fl) звезды: V388 Cas (Er), V1589 Cyg (Fl/Er), GJ 1078 (Fl), GL 851.1, V780 Tau (Fl/Er) и V962 Tau (Er).

В статье (Синицына и др., 2020) авторами представлена идентификация собственных наблюдений в рентгеновском диапазоне спектра с объектами, указанными в таблице 1. К сожалению, авторы не совместили полученные данные, представленные в виде изофот, с оптическими изображениями изучаемых областей и не выполнили идентификацию возможных источников поля.

Таблица 1.

| Объект | Рентгеновский источник | Объект | Рентгеновский источник |
|--------|------------------------|--------|------------------------|
| V388 Cas | 1RXS J010318.0+622146 | GL 851.1 | - |
| V1589 Cyg | 2RSX J204249.0+412246 | V780 Tau | 2RXS J054025.1+244839 |
| GJ 1078 | 1RXS J052327.7+222649 | V962 Tau | 2RXS J054552.1+225248 |

*Оптические отождествления, рентгеновские источники и возможные кандидаты для ТэВ излучения*

Для проведения отождествления в оптическом диапазоне спектра изображений источников ТэВ излучения применялся интерактивный атлас неба Aladin (Боннарель и др., 2000) с подгружаемыми базами данных, поддерживаемыми Центром астрономических данных в Страсбурге.

Основные данные об отождествляемых звёздах представлены в таблице 2.

Таблица 2.

| Объект | SIMBAD | R.A. (2000) | Decl. (2000) | pmRA (mas) | pmDE (mas) | Type | B (mag) | V (mag) | Sp. type | Dist. |
|--------|--------|-------------|--------------|------------|------------|------|---------|---------|----------|-------|
| V388 Cas | Wolf 47 | 01 03 19.83 | +62 21 55.8 | 730.74 | 86.352 | Er | 15.46 | 13.78 | M5V | 15" |
| V1589 Cyg | V* V1589 Cyg | 20 42 49.16 | +41 23 00.0 | 67.3 | -31.1 | Fl/Er | 15.29 | 13.59 | - | 15" |
| GJ 1078 | G 85-69 | 05 23 49.05 | +22 32 38.8 | 236.0 | -300.0 | Fl | 17.35 | 15.52 | M4.5V | 9" |
| GL 851.1 | BD+30 4633 | 22 12 06.42 | +31 33 41.1 | -344 | -418.8 | Fl | 11.43 | 10.15 | K5V/dM0e | 21" |
| V780 Tau | G 100-28 | 05 40 25.73 | +24 48 07.8 | 107.0 | -376.0 | Fl/Er | 16.74 | 14.94 | M7V | 10" |
| V962 Tau | V* V962 Tau | 05 45 51.94 | +22 52 47.4 | 0.142 | -9.959 | Er | 13.5 | - | - | 10" |

В первой колонке таблицы приведено обозначение звезды по (Синицына и др., 2020), во второй – основное обозначение по SIMBAD. Третья и четвёртая колонки содержат экваториальные координаты объектов на эпоху 2000 года. В пятой и шестой – собственные движения звёзд по прямому восхождению и склонению в угловых миллисекундах в год. В седьмой – тип переменности звезды. В восьмой и девятой – блеск в полосах B и V. Десятая колонка содержит информацию о спектральном типе звезды, и одиннадцатая – угловое расстояние от максимума потока в диапазоне гамма излучения до оптического изображения звезды.

*V388 Cas*

Максимум потока в диапазоне гамма излучения (МГИ) находится на расстоянии 15 угловых секунд от оптического положения V388 Cas (рис. 1). На рисунке максимум соответствует наиболее красному оттенку изофоты и практически совпадает с красным маркером и окружностью того же цвета, очевидно ограничивающей область ошибок (error radius - ER) в определении координат рентгеновского источника (РИ), которым авторы (Синицына и др., 2020) отметили положение 1RXS J010318.0+622146. Однако, реально, положение источника находится несколько ниже (зелёные окружности меньшего радиуса – 1×ER и большего радиуса – 3×ER, отмечены цифрами 2 и 3) на расстоянии 23".

Красной окружностью с ER = 0".8 (цифра – 4) – указано положение РИ 4XMM J010321.3+622157 (Уэбб и др., 2020), зарегистрированного телескопами XMM-Newton, и находящегося на расстоянии 28" от МГИ. Желтой окружностью с ER = 3".9, с удалением на 27" (цифра – 5), – указано положение РИ 2SXPS J010321.2+622159 (Эванс и др., 2020), наблюдавшегося обсерваторией Swift.

Малыми окружностями желтого цвета на рисунке показаны положения звёзд из каталога GAIA DR2 (GAIA, 2018). Цифра 1 соответствует V388 Cas. Разница в положении звезды по данным GAIA и на оптическом изображении обусловлена значительным собственным движением (730.740 по прямому восхождению и 86.352 по склонению угловых миллисекунд в год, соответственно).

Обращают на себя внимание несколько областей в окрестностях V388 Cas, в которых расположены слабые источники гамма-излучения, ограниченные прямоугольниками желтого цвета (помечены цифрами 6, 7, 8, 9 и 10). Внутри находятся объекты из каталога GAIA DR2 со следующими характеристиками (таблица 3).

Таблица 3.

| № | R.A. (2000) | Decl. (2000) | Plx | pmRA | pmDE | G | $G_{BP}$ | $G_{RP}$ | Teff | Rad | Lum |
|---|---|---|---|---|---|---|---|---|---|---|---|
| 1 | 015.83942594 | +62.36588063 | 101.6371 | 730.740 | 86.352 | 11.9238 | 13.9068 | 10.6032 | 3346.00 | - | - |
| 6 | 015.81867355 | +62.36274707 | 0.4845 | -1.450 | -0.828 | 17.8200 | 18.5362 | 16.9453 | - | - | - |
| 7 | 015.79997556 | +62.36915143 | -0.0123 | 0.921 | 0.653 | 19.6506 | 20.5764 | 18.5647 | - | - | - |
| 8 | 015.81488911 | +62.37208876 | 0.3020 | -1.116 | -0.125 | 16.5141 | 17.8314 | 15.3511 | 3670.99 | - | - |
| 9↑ | 015.82903889 | +62.37352009 | 1.5970 | 2.231 | -7.581 | 15.1327 | 15.7619 | 14.3744 | 4547.58 | 0.92 | 0.326 |
| 9↓ | 015.82802289 | +62.37205461 | 0.6506 | 8.320 | -2.823 | 16.7946 | 17.5051 | 15.9437 | 4312.85 | 1.23 | 0.470 |
| 10 | 015.83925097 | +62.36866398 | 0.3280 | -3.790 | -0.808 | 13.5627 | 14.4896 | 12.6116 | 3945.89 | 14.34 | 44.925 |

В таблице 3 указаны номера прямоугольников со слабыми источниками гамма излучения, в которых обнаружены объекты, за исключением № 1, соответствующего V388 Cas. Номера 9↑ и 9↓ принадлежат верхнему и нижнему объектам в девятом прямоугольнике. Далее в столбцах: координаты объектов, параллакс, собственное движение по прямому восхождению и склонению по каталогу GAIA DR2, средние звёздные величины в полосах G, $G_{BP}$ и $G_{RP}$ в фотометрической системе GAIA, эффективная температура, радиусы и светимости звёзд в сравнении с солнечными.

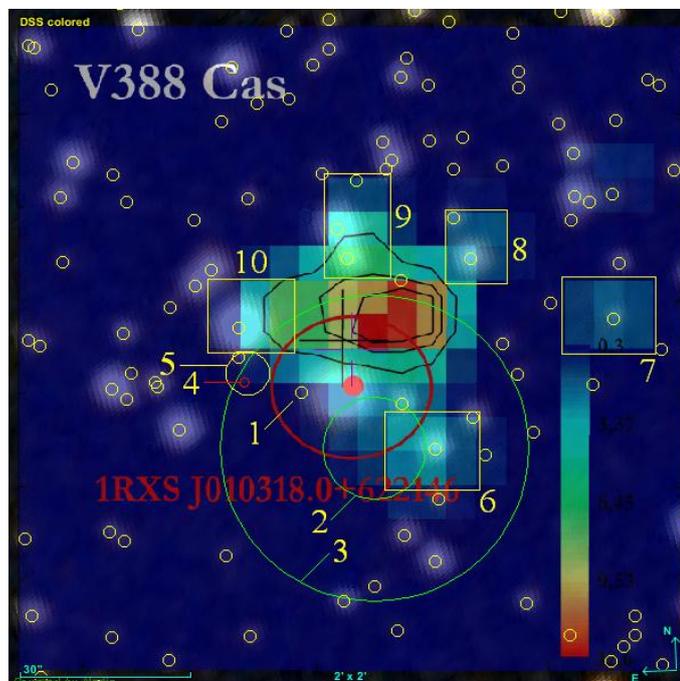

Рис. 1. Комбинированное оптическое (DSS colored) и ТэВ изображения области V388 Cas. Окружностями обозначены положения и ER рентгеновских источников, прямоугольниками – возможные оптические отождествления. Малыми жёлтыми окружностями отмечены звёзды из каталога GAIA DR2.

Отметим, что согласно данным 2-го каталога источников ROSAT (Болир и др., 2016), объект 1RXS J010318.0+622146 ассоциируется с рентгеновским источником 2RXS J010318.3+622140 и демонстрирует слабую переменность (рис. 2).

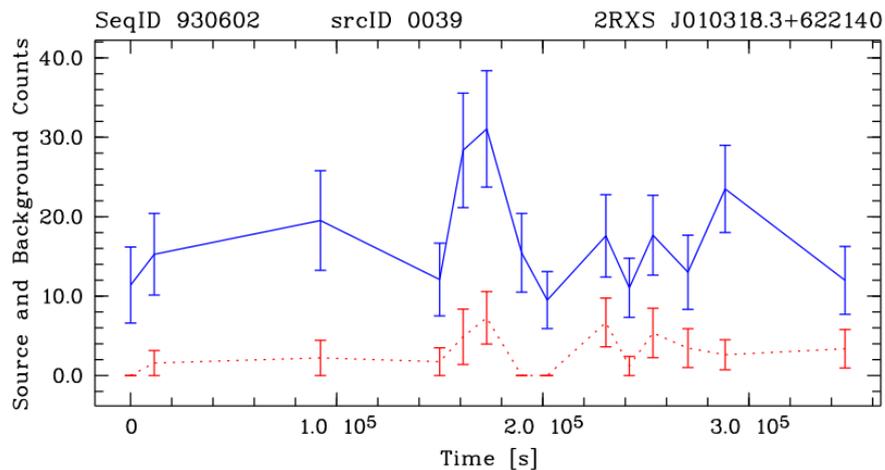

Рис. 2. Кривая блеска объекта 2RXS J010318.3+622140 в рентгеновском диапазоне спектра. По осям отложены: время от начала наблюдений в секундах и отсчеты детектора для источника (синяя кривая) и фона (красная кривая). Указаны ошибки измерений.

Следует также отметить, что РИ 2SXPS J010321.2+622159 имеет более близкое расположение к объекту 10 из таблицы 3, чем к V388 Cas. А в случае использования критерия 3×ER, однозначно попадает в поле рентгеновского источника.

*V1589 Cyg*

Координаты, приведенные в статье Синициной и др. на 1 градус (рис. 3а) отличаются от координат объекта в базе данных SIMBAD (Венга и др., 2000). Звезда находится в верхней части изображения, площадка зарегистрированного гамма излучения в нижней части рисунка. При приведении координат ТэВ изображения в соответствие с координатами V1589 Cyg, максимум потока в гамма диапазоне находится на расстоянии ~15 угловых секунд от оптического положения звезды (рис. 3б).

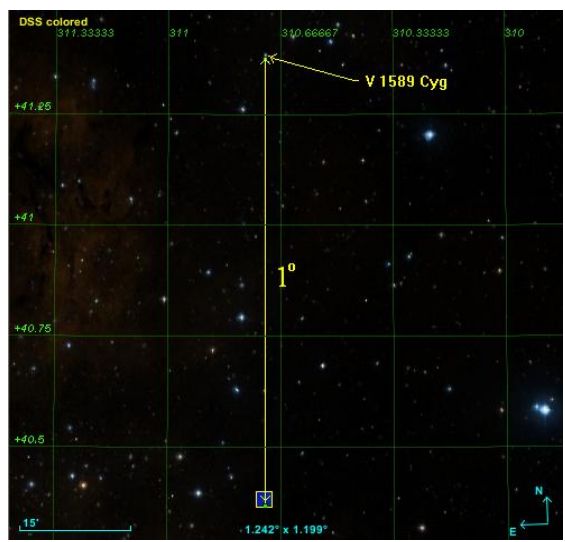

Рис. 3а. Реальное положение V1589 Cyg (в верхней части рисунка), и положение ТэВ источника, в соответствие с координатами из (Синицына и др., 2020).

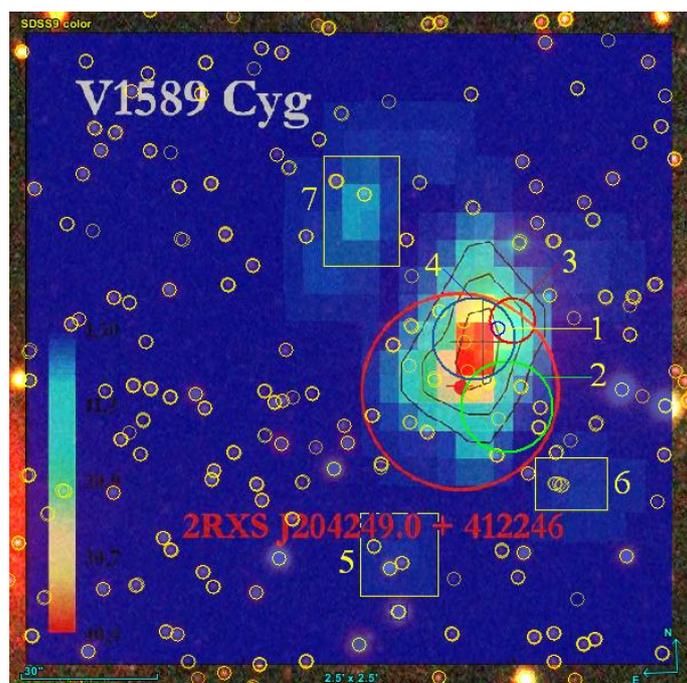

Рис. 3б. Комбинированное оптическое (SDSS9 color) и ТэВ изображения области V1589 Cyg. Окружностями обозначены положения и ER рентгеновских источников, прямоугольниками области с возможными оптическими отождествлениями. Малыми желтыми окружностями отмечены звёзды из каталога SDSS DR12.

Цифрами на рисунке 3б обозначены: 1 – положение V1589 Cyg, согласно координат на эпоху 2000 года (синяя окружность малого радиуса); 2 – зелёная окружность, соответствующая 1×ER источнику 1RXS J204249.0+412242; 3 - красная окружность малого радиуса, соответствует 1×ER источника XMMSL2 J204248.9+412302 (XMM-SSC, 2017); 4 - синяя окружность – область ошибок положения объекта 1SWXRT J204249.6+412257 из каталога обсерватории Swift (Д'Элия и др., 2013). Малыми желтыми окружностями отмечены звёзды из каталога SDSS DR12 (Алям и др., 2015).

Большая окружность красного цвета – положение и ER источника 2RXS J204249.0+412245 из (Синицына и др., 2020). Отметим, что более близкими по координатам к V1589 Cyg, и имеющими меньший ER, являются источники рентгеновского излучения XMMSL2 J204248.9+412302 и 1SWXRT J204249.6+412257.

Представляют интерес объекты (рис. 4), расположенные в областях слабого гамма излучения, обозначенные на рисунке 3б желтыми прямоугольниками с цифрами 5, 6 и 7. Максимум потока в них превышает уровень фона более чем в 3 раза для областей 5 и 6, и около 7 раз для области 7 (Оценка потока производилась по шкале интенсивностей изофоты, приведенной в левой части рисунка). Некоторые из объектов в данных областях (таблица 4) могут являться источниками гамма излучения.

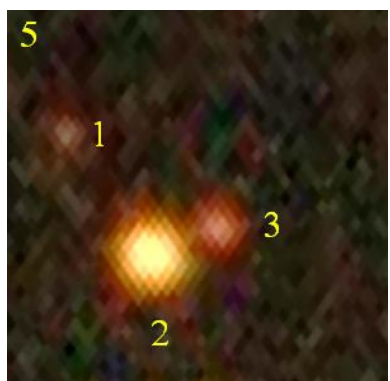

Рис. 4а. Карта области 5.

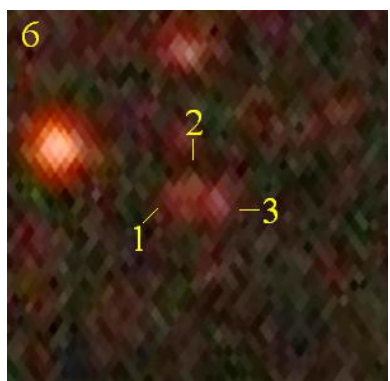

Рис. 4б. Карта области 6.

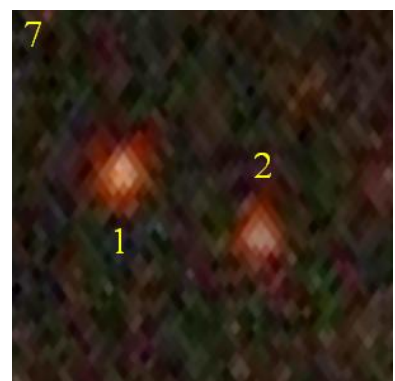

Рис. 4в. Карта области 7.

Таблица 4

| № | R.A. (2000) | Decl. (2000) | n | u | er_u | g | er_g | r | er_r | i | er_i | z | er_z | Sp |
|---|---|---|---|---|---|---|---|---|---|---|---|---|---|---|
| 5.1 | 310.71493530 | +41.36992645 | 3 | 24.966 | 0.836 | 25.037 | 0.585 | 23.865 | 0.474 | 22.070 | 0.161 | 20.869 | 0.227 | K5 |
| 5.2 | 310.71368408 | +41.36857986 | 4 | 25.075 | 0.905 | 21.431 | 0.052 | 19.367 | 0.014 | 18.330 | 0.010 | 17.610 | 0.018 | M5 |
| 5.3 | 310.71267700 | +41.36889267 | 3 | 24.885 | 0.898 | 24.976 | 0.548 | 23.218 | 0.312 | 21.014 | 0.068 | 19.570 | 0.077 | M3 |
| 6.1 | 310.70025635 | +41.37376785 | 2 | 24.338 | 0.873 | 24.603 | 0.590 | 23.691 | 0.572 | 21.437 | 0.131 | 20.042 | 0.160 | K2 |
| 6.2 | 310.69989600 | +41.37380900 | 1 | 25.673 | 0.884 | 26.649 | 0.382 | 25.641 | 0.626 | 23.606 | 0.801 | 20.507 | 0.256 | K5 |
| 6.3 | 310.69964600 | +41.37366486 | 2 | 24.700 | 0.937 | 24.490 | 0.482 | 24.584 | 0.618 | 22.528 | 0.273 | 20.879 | 0.231 | A2 |
| 7.1 | 310.71801758 | +41.39233780 | 4 | 25.696 | 0.720 | 24.912 | 0.580 | 22.309 | 0.154 | 20.643 | 0.055 | 19.247 | 0.067 | M6 |
| 7.2 | 310.71572876 | +41.39153671 | 3 | 24.945 | 0.779 | 24.820 | 0.564 | 22.899 | 0.233 | 21.180 | 0.075 | 19.726 | 0.087 | M4 |

В таблице 4 указаны номера прямоугольников со слабыми источниками гамма-излучения и соответствующие им номера объектов в этих областях. Далее в столбцах: координаты объектов, число наблюдений, средние значения звёздных величин и ошибок их определения в полосах u, g, r, i, z в цветовой системе каталога SDSS DR12. В последней колонке – ориентировочный усреднённый спектральный тип объектов, найденный по показателям цвета (Пеко и Мамежек, 2013, Мамежек, 2019).

Все объекты в областях, за исключением 6.1 классифицированы в каталоге SDSS DR12 как звёзды и имеют хорошее качество наблюдений. Объект 6.1 имеет две классификации, как звезда и как галактика на основе моделирования фотометрического профиля его изображения. Возможно, все три объекта области 6, учитывая их близкое расположение, являются единым протяжённым объектом – галактикой, а классификация объектов 2 и 3 – ошибочной.

*GJ 1078*

Рентгеновский источник 1RXS J052327.7+222649 (Водис и др., 2000), представленный в статье (Синицына и др., 2020), как возможная идентификация для GJ 1078 находится на расстоянии более 7 угловых минут от максимума ТэВ изображения и имеет радиус ошибок значительно меньший, чем приведен в публикации. На рисунке 5 положение 1RXS J052327.7+222649 и ER (12"), соответствующий ему, (окружность зелёного цвета), показан в правой нижней части изображения.

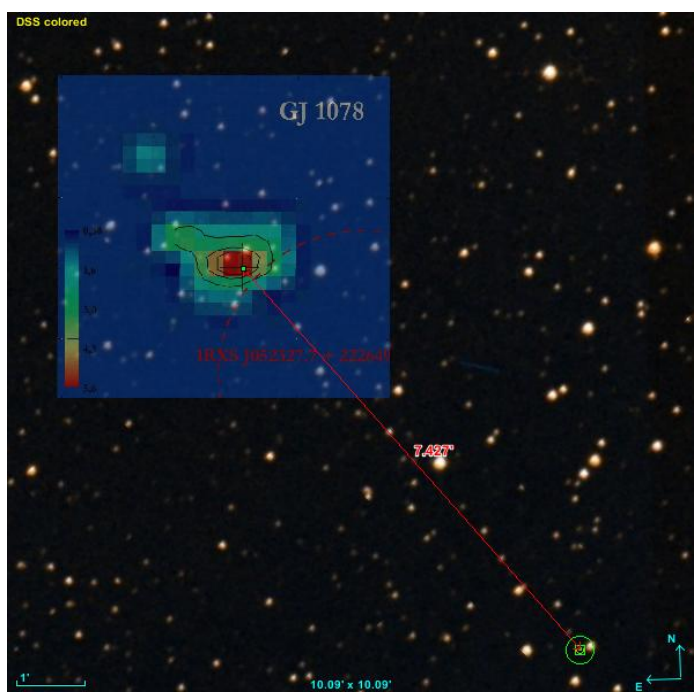

Рис. 5. Комбинированное оптическое (DSS colored) и
ТэВ изображения области GJ 1078.

Возможной идентификацией для РИ 1RXS J052327.7+222649 является звезда, попадающая в ER, GAIA DR2 3414565834307772672 с характеристиками, указанными в таблице 5.

Таблица 5.

| R.A. (2000) | Decl. (2000) | Plx | pmRA | pmDE | G | $G_{BP}$ | $G_{RP}$ | Teff |
|---|---|---|---|---|---|---|---|---|
| 080.86267583 | +22.44798972 | 10.0892 | -21.366 | -28.036 | 14.3127 | 15.6859 | 13.1702 | 3887.82 |

В таблице приведены координаты звезды, её параллакс, собственное движение по прямому восхождению и склонению в стандартных единицах, звёздные величины в полосах G, $G_{BP}$, $G_{RP}$ в фотометрической системе каталога GAIA DR2, и эффективная температура (K).

Объект 1RXS J052327.7+222649, занесенный в каталог источников ROSAT (Болир и др., 2016), как 2RXS J052327.6+222649 демонстрирует незначительную переменность на уровне фона в рентгеновском диапазоне, показанную на рисунке 6.

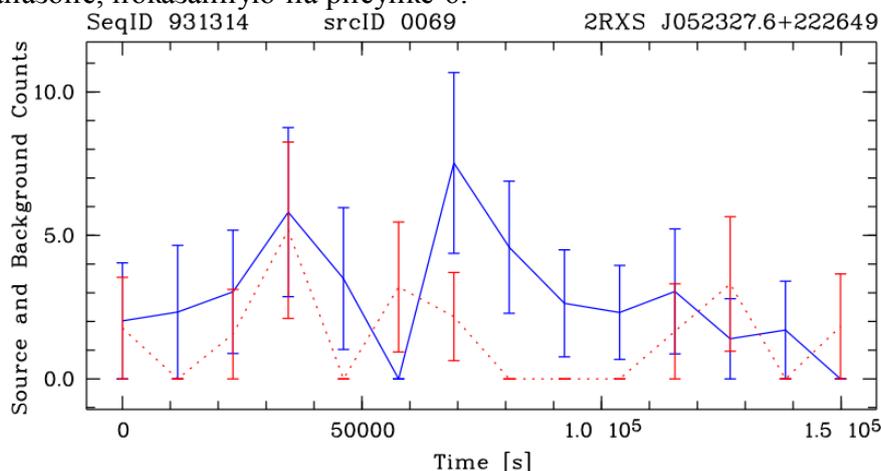

Рис. 6. Кривая блеска объекта 2RXS J052327.6+222649 в рентгеновском диапазоне спектра. По осям отложены: время от начала наблюдений в секундах и отсчеты детектора для источника (синяя кривая) и фона (красная кривая). Указаны ошибки измерений.

Рассмотрим детально область ТэВ изображения для GJ 1078 из (Синицына и др., 2020), представленную на рисунке 7. Цифрами обозначено: 1 – положение GJ 1078 из каталога GAIA DR2. Звезда находится на расстоянии 9" от максимума ТэВ источника; 2 – РИ 1RXH J052349.0+223251 (RST, 2000), имеющий ER 9" (зелёная окружность); 3 – РИ 1RXH J052349.5+223236 (RST, 2000) с ER 12" (желтая окружность), который наилучшим образом согласуется по положению с GJ 1078; 4, 5 и 6 – области повышенного гамма излучения (рис. 8а, б, в), объекты из которых содержатся в таблице 6. Отметим, также, что в область РИ 1RXH J052349.5+223236 попадает звезда, отмеченная в таблице 6 под номером 7.

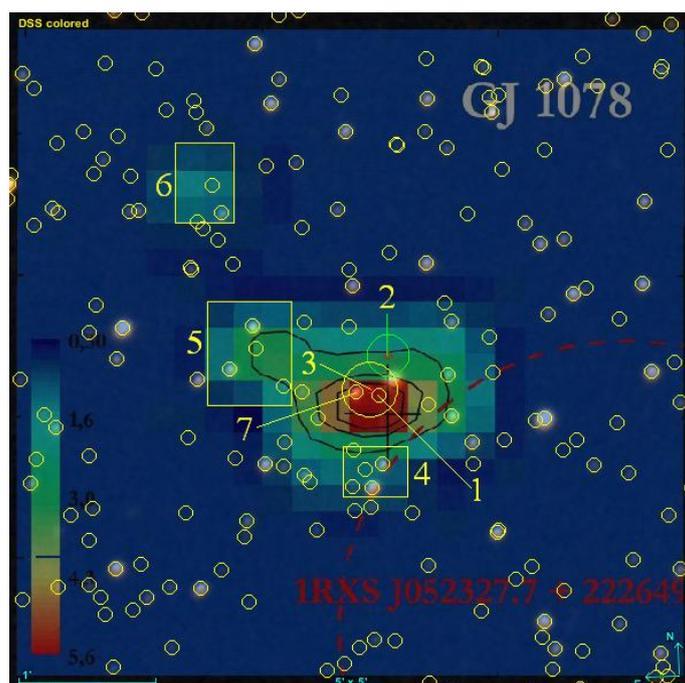

Рис. 7. Комбинированное оптическое (DSS colored) и ТэВ изображения области GJ 1078 с указанием объектов, и областей для которых выполнено отождествление.

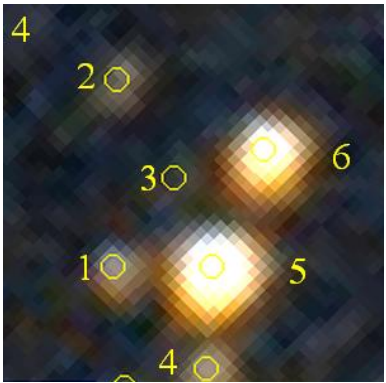 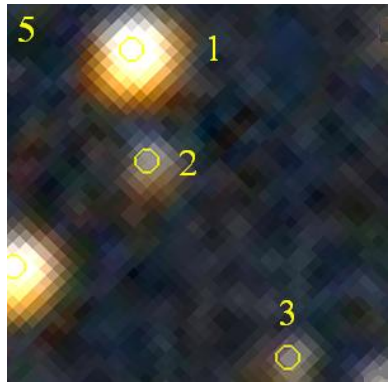 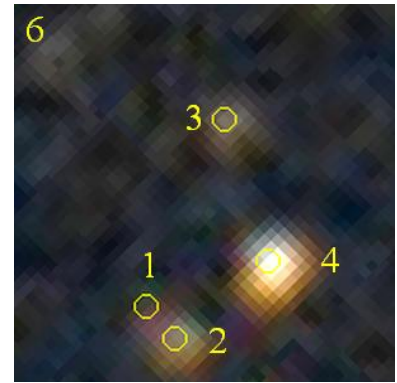

Рис. 8а. Карта области 4.   Рис. 8б. Карта области 5.   Рис. 8в. Карта области 6.

Таблица 6

| №   | R.A. (2000)   | Decl. (2000)  | Plx     | pmRA   | pmDE    | G       | $G_{BP}$ | $G_{RP}$ | Teff    | Rad  | Lum   |
|-----|---------------|---------------|---------|--------|---------|---------|----------|----------|---------|------|-------|
| 4.1 | 080.95877765  | +22.53166155  | 0.4863  | -0.857 | -2.930  | 19.2419 | 20.3034  | 18.1495  | -       | -    | -     |
| 4.2 | 080.95869733  | +22.53614840  | 1.6889  | -1.487 | 0.121   | 19.8610 | 20.4998  | 18.8138  | -       | -    | -     |
| 4.3 | 080.95720994  | +22.53378669  | 0.8952  | 5.567  | -10.956 | 20.4132 | 21.0040  | 19.0972  | -       | -    | -     |
| 4.4 | 080.95637812  | +22.52918229  | 0.4649  | -0.462 | -2.973  | 19.1219 | 20.0388  | 18.1138  | -       | -    | -     |
| 4.5 | 080.95624474  | +22.53162562  | 0.6151  | -1.024 | -2.809  | 15.6819 | 16.4185  | 14.8271  | 4430.25 | 2.00 | 1.388 |
| 4.6 | 080.95490905  | +22.53444779  | 0.6055  | 0.384  | -8.921  | 16.7066 | 17.6193  | 15.7507  | 3883.25 | -    | -     |
| 5.1 | 080.97186813  | +22.55128457  | 0.8428  | 2.418  | -3.582  | 16.2663 | 17.0784  | 15.3653  | 4008.50 | 1.53 | 0.543 |
| 5.2 | 080.97147333  | +22.54862508  | 0.1456  | 0.231  | -1.259  | 19.3775 | 20.1277  | 18.4256  | -       | -    | -     |
| 5.3 | 080.96782341  | +22.54390129  | -0.0681 | -0.261 | -0.314  | 19.5618 | 20.4901  | 18.5803  | -       | -    | -     |
| 6.1 | 080.97919404  | +22.56397196  | 0.5140  | 1.560  | -5.213  | 20.5542 | 20.9304  | 19.2965  | -       | -    | -     |
| 6.2 | 080.97843061  | +22.56320793  | 0.7603  | 0.791  | -0.843  | 19.4131 | 20.0881  | 18.3599  | -       | -    | -     |
| 6.3 | 080.97714800  | +22.56847806  | 0.0570  | 1.022  | -0.240  | 20.0700 | 20.9232  | 18.9711  | -       | -    | -     |
| 6.4 | 080.97603549  | +22.56508820  | 0.1715  | 1.641  | -1.392  | 17.6060 | 18.5273  | 16.6349  | -       | -    | -     |
| 7   | 080.95844612  | +22.54333320  | 0.4040  | -1.151 | -2.167  | 16.7120 | 17.4675  | 15.8276  | 4239.00 | -    | -     |

Описание колонок в таблице 6 соответствует описанию к таблице 3.

Из рассматриваемых звёзд, попадающих в области гамма излучения, превышающие уровень фона, представляют интерес объекты 4.5 (K6.5V) и 5.1 (K8V), для которых имеется информация по эффективной температуре, радиусу и светимости. Усреднённый спектральный тип объектов, определённый по информации из каталога GAIA DR2 о показателях цвета $G - G_{RP}$ и $G_{BP} - G_{RP}$, дал следующие результаты: для 4.5 – K6.5V и 5.1 – K8V.

### GL 851.1

Карта для данного объекта, представленная в статьях (Синицына и др., 2019, Синицына и др., 2020) является зеркальной. Поэтому для наложения её на оптическое изображение она была развёрнута относительно вертикальной оси на 180º (рис. 9). Желтые окружности – объекты из каталога GAIA DR2. Цифрой 1 обозначена звезда GL 851.1. Её расстояние до максимума гамма излучения составляет 21 угловую секунду. Данные для объектов 2 и 3 из каталога GAIA DR2 содержатся в таблице 7. Описание колонок в таблице 7 соответствует описанию к таблице 3.

Таблица 7

| № | R.A. (2000)  | Decl. (2000) | Plx    | pmRA   | pmDE   | G       | $G_{BP}$ | $G_{RP}$ | Teff | Rad | Lum |
|---|--------------|--------------|--------|--------|--------|---------|----------|----------|------|-----|-----|
| 2 | 333.03401553 | +31.55554772 | 1.2811 | -0.320 | 4.852  | 19.7206 | 20.2451  | 18.4217  | -    | -   | -   |
| 3 | 333.03330817 | +31.55555801 | 0.1898 | 0.057  | -5.447 | 17.6793 | 18.0652  | 17.0552  | -    | -   | -   |

Объекты 2 и 3 находятся на расстоянии 2.3 угловые секунды друг от друга и попадают в область слабого гамма излучения, связываемого с GL 851.1. Интерес к ним обусловлен тем, что объект 3 является карликом, согласно (Стассун и др., 2019), с усреднённым спектральным классом G2 по данным фотометрии SDSS DR16 (Ахумада и др., 2020). Объект 2 по наблюдениям GAIA также является звездой. Для неё определён параллакс и собственные движения. Однако, в случае определения собственного движения имеют место значительные ошибки. Для pmRA = -0.320 угловых миллисекунд в год ошибка составляет 0.589, а для pmDE = 4.852, соответственно, 0.821. Ошибки в определении параллакса также значительны: 0.4273 угловые миллисекунды. При использовании критерия 3σ величина ошибок незначительно превышает значение определённого параллакса (3σ = 1.2819).

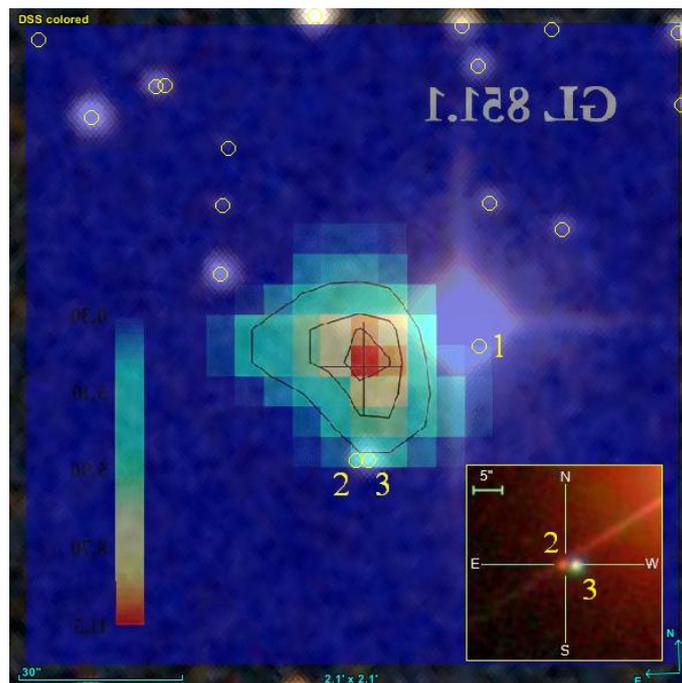

Рис. 9. Комбинированное оптическое (DSS colored) и ТэВ изображения области GL 851.1. На врезке – объекты 2 и 3 в атласе SDSS.

По данным SDSS DR16 объект 2 является галактикой с фотометрическим (Аяйдзу и др., 2008) красным смещением 0.171 ± 0.1692. Значительные ошибки в определении красного смещения ставят его достоверность под сомнение. Возможно, и это хорошо видно на врезке рисунка 9, определение объекта 2 как галактики обусловлено проекцией дифракционного луча от GL 851.1 и последующей некорректной обработкой изображения.

*V780 Tau*

Максимум потока в диапазоне гамма излучения находится на расстоянии ~ 10 угловых секунд от оптического положения V780 Tau, обозначенного цифрой 1 на рисунке 10. Все рассматриваемые ниже объекты из каталога GAIA DR2 прорисованы желтыми окружностями. Учитывая значительное собственное движение V780 Tau, на оптическом изображении области положение её занимает звезда № 3. Сама же V780 Tau на эпоху 2000.0 сместилась ниже и левее.

Данные для объектов, попадающих в области повышенного гамма излучения (обозначены желтыми прямоугольниками) в окрестностях V780 Tau, представлены в таблице 8. Описание колонок в таблице 3 соответствует таблице 8.

Таблица 8

| № | R.A. (2000) | Decl. (2000) | Plx | pmRA | pmDE | G | $G_{BP}$ | $G_{RP}$ | Teff | Rad | Lum |
|---|---|---|---|---|---|---|---|---|---|---|---|
| 2 | 085.11024188 | +24.80279535 | 0.2364 | 3.406 | -1.215 | 16.0450 | 16.6976 | 15.2211 | 4332.67 | - | - |
| 3 | 085.10686509 | +24.80357145 | 0.0514 | 1.992 | -0.753 | 19.4176 | 20.1994 | 18.4724 | - | - | - |
| 4 | 085.10766325 | +24.79806903 | - | - | - | 20.5223 | 21.0957 | 19.0595 | - | - | - |
| 5 | 085.10796446 | +24.79425983 | 0.0066 | 2.105 | -0.751 | 18.7862 | 19.4006 | 17.9025 | - | - | - |
| 6 | 085.10315316 | +24.79172505 | -0.0479 | 1.568 | -0.873 | 18.5154 | 19.0813 | 17.5660 | - | - | - |

Под V780 Tau цифрой 6 обозначен объект из GAIA DR2 (GAIA, 2018) с $G_{RP}$ звёздной величиной $17^m.566$, который совпадает по положению со слабым регистрируемым потоком гамма-излучения (обозначено желтым прямоугольником). Звезда имеет координаты $05^h40^m24^s.76$ +24°47′30″.2, для неё определён параллакс и собственное движение. Однако данные каталога IGAPS (Монгуё и др., 2020), полученные на основе наблюдений на телескопе Исаака Ньютона (INT) между 2003 и 2018 годами, вносят неопределённость в классификацию объекта как звезды или галактики.

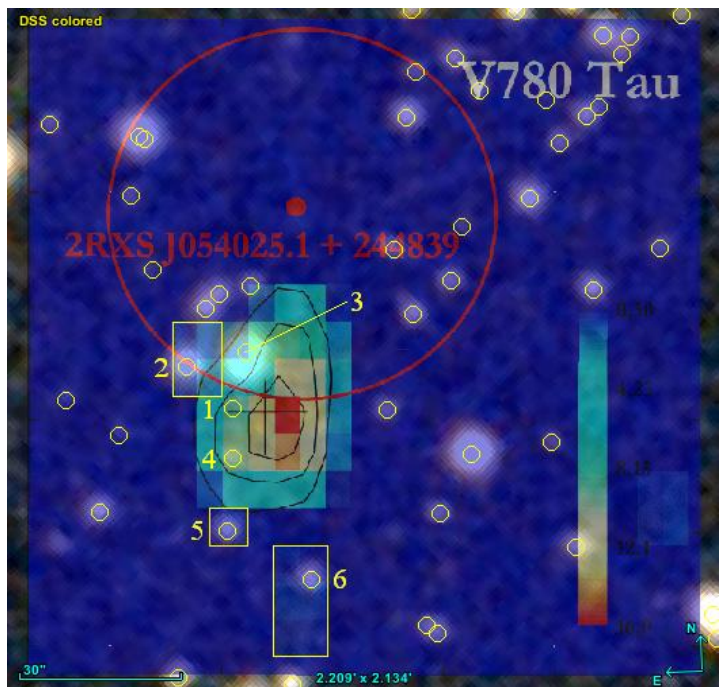

Рис. 10. Наложение изофоты гамма излучения области V780 Tau на оптическое изображение (DSS colored).

Отметим, что V780 Tau демонстрирует слабую переменность в рентгеновском диапазоне спектра по данным 2-го каталога источников ROSAT (Болир и др., 2016). Изменение регистрируемого потока 2RXS J054025.1+244839 с течением времени продемонстрировано на рисунке 11.

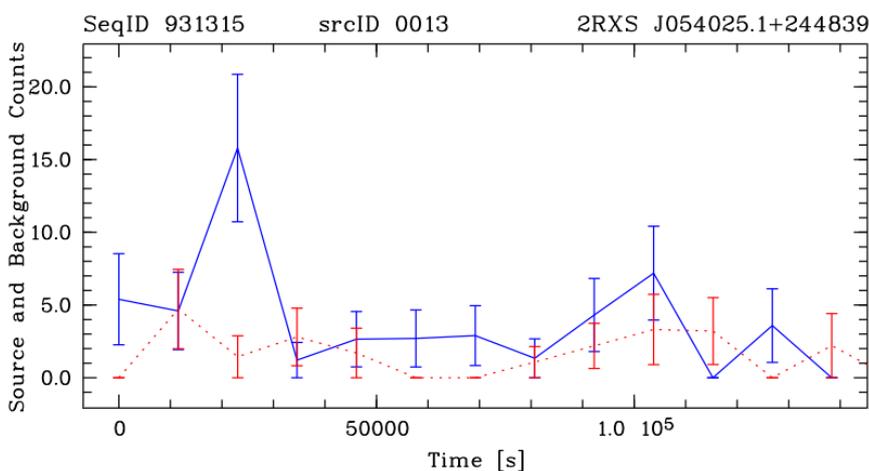

Рис. 11. Кривая блеска объекта 2RXS J054025.1+244839 = V780 Tau в рентгеновском диапазоне спектра. По осям отложены: время от начала наблюдений в секундах и отсчеты детектора для источника (синяя кривая) и фона (красная кривая). Указаны ошибки измерений.

*V962 Tau*

V962 Tau, обозначенная на рисунке 12 цифрой 1 находится на расстоянии ~10 угловых секунд от максимума потока в диапазоне гамма излучения. Данные об объектах, попадающих в область более слабого гамма излучения, представлены в таблице 9. Описание колонок в таблице 9 соответствует описанию к таблице 3.

Таблица 9

| № | R.A. (2000) | Decl. (2000) | Plx | pmRA | pmDE | G | $G_{BP}$ | $G_{RP}$ | Teff | Rad | Lum |
|---|---|---|---|---|---|---|---|---|---|---|---|
| 2 | 086.46128416 | +22.88003290 | 8.9771 | 3.820 | -37.943 | 12.9985 | 14.4091 | 11.8235 | 4078.39 | 0.61 | 0.093 |
| 3 | 086.46016711 | +22.87969449 | 1.1031 | 2.352 | -5.134 | 18.6179 | 19.0753 | 17.0258 | - | - | - |

Представляет интерес звезда № 2 вблизи V962 Tau. В пределах одной угловой секунды по координатам положение звезды из каталога GAIA DR2 совпадает с рентгеновским источником 2SXPS J054550.7+225249 из "Чистой" версии каталога точечных источников рентгеновского телескопа обсерватории Swift (Эванс и др., 2020). Таким образом, GAIA DR2 3427527766435285248 является возможным оптическим кандидатом на идентификацию с РИ 2SXPS J054550.7+225249.

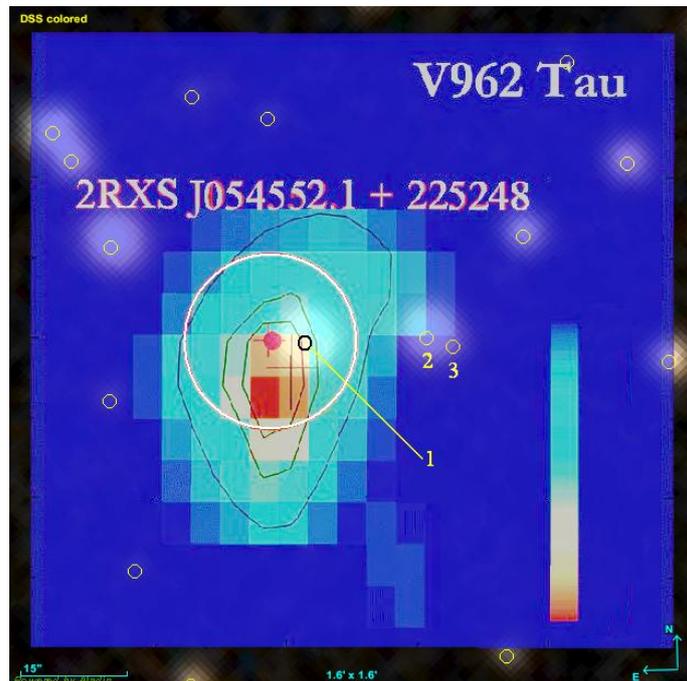

Рис. 12. Локализация ТэВ источника на фоне оптического изображения области V962 Tau (DSS colored).

V962 Tau демонстрирует значительную переменность в рентгеновском диапазоне на уровне фона (с подобием квазипериодичности) по данным (Болир и др., 2016), что проиллюстрировано на рисунке 13.

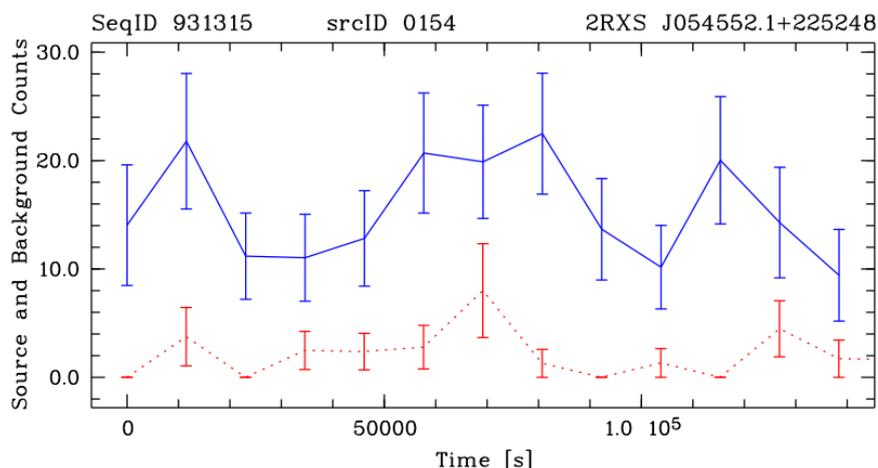

Рис. 13. Кривая блеска объекта 2RXS J054552.1+225248 = V962Tau в рентгеновском диапазоне спектра. По осям отложены: время от начала наблюдений в секундах и отсчеты детектора для источника (синяя кривая) и фона (красная кривая). Указаны ошибки измерений.

**Заключение**

В результате проведённого анализа было установлено соответствие источников ТэВ излучения красным карликовым звёздам. Высказано предположение о возможной идентификации в оптической области спектра других объектов поля. В частности рентгеновских источников: 2SXPS J010321.2+622159 = 522864272037651712, 1RXS J052327.7+222649 = 3414565834307772672, 2SXPS J054550.7+225249 = 3427527766435285248. После знака равенства указано уникальное для

всех выпусков данных GAIA обозначение источника. Формально, перед обозначением должно стоять «Gaia DR2».

Составлен список 39 объектов, попадающих в области слабого гамма излучения, превышающего уровень фона, и в окрестностях звёзд, рассмотренных в статьях (Синицына и др., 2019, Синицына и др., 2020). Следует, также отметить, что координатное обеспечение рисунков с распределением гамма излучения вблизи рассмотренных красных карликов имеет ошибки. Это несколько усложнило идентификацию в оптическом диапазоне.

## Литература